\documentclass{nature}

\usepackage{graphicx}

\usepackage{epsfig}

\title{Large intrinsic anomalous Hall effect in half-metallic ferromagnet Co$_{3}$Sn$_{2}$S$_{2}$ with magnetic Weyl fermions}

\author{Qi Wang,$^{1,\dag}$ Yuanfeng Xu,$^{2,\dag}$ Rui Lou,$^{1,\dag}$ Zhonghao Liu,$^{3}$ Man Li,$^{1,4}$ Yaobo Huang,$^{4}$ Dawei Shen,$^{3}$ Hongming Weng,$^{2,5,*}$ Shancai Wang,$^{1,*}$ and Hechang Lei$^{1,*}$}

\begin{document}

\maketitle

\begin{affiliations}
 \item Department of Physics and Beijing Key Laboratory of Opto-electronic Functional Materials $\&$ Micro-nano Devices, Renmin University of China, Beijing 100872, China
 \item Beijing National Laboratory for Condensed Matter Physics and Institute of Physics, Chinese Academy of Sciences, Beijing 100190, China
 \item State Key Laboratory of Functional Materials for Informatics and Center for Excellence in Superconducting Eletronics, SIMIT, Chinese Academy of Sciences, Shanghai 200050, China
 \item Shanghai Synchrotron Radiation Facility, Shanghai Institute of Applied Physics, Chinese Academy of Sciences, Shanghai 201204, China
 \item Collaborative Innovation Center of Quantum Matter, Beijing, China
 \end{affiliations}

\leftline{$^\dag$These authors contributed equally to this work.}
\leftline{$^*$Corresponding authors: hmweng@iphy.ac.cn, scw@ruc.edu.cn, hlei@ruc.edu.cn}\vspace*{1cm}

\begin{abstract}
The origin of anomalous Hall effect (AHE) in magnetic materials is one of the most intriguing aspect in condensed matter physics and has been controversial for a long time. Recent studies indicate that the intrinsic AHE is closely related to the Berry curvature of occupied electronic states. In a magnetic Weyl semimetal with broken time-reversal symmetry, there are significant contributions on Berry curvature around Weyl nodes, which would lead to a large intrinsic AHE. Here, we report the large intrinsic AHE in the half-metallic ferromagnet Co$_{3}$Sn$_{2}$S$_{2}$ single crystal. By systematically mapping out the electronic structure of Co$_{3}$Sn$_{2}$S$_{2}$ theoretically and experimentally, the large intrinsic AHE should originate from the Weyl fermions near the Fermi energy. Furthermore, the intrinsic anomalous Hall conductivity depends linearly on the magnetization and this can be attributed to the sharp decrease of magnetization and the change of topological characteristics.
\\
\end{abstract}

The ordinary Hall effect that arises from the Lorentz force deflecting the moving charge carriers has been well understood.\cite{Hall1} In contrast, the anomalous Hall effect (AHE) \cite{Hall2} has attracted tremendous interests because of the fundamental physics and great potential in technical application,\cite{LBerger,Hurd1} but the microscopic origin of AHE is still not fully solved. The key issue is whether the effect is intrinsic or extrinsic. It is now recognized that there are three mechanisms to account for the AHE.\cite{Nagaosa} One is the extrinsic mechanism related to the scattering affected by the spin-orbit interaction, i.e., the skew scattering and side jump mechanisms.\cite{Smit,Berger} The skew scattering model predicts that the $\rho_{xy}^{A}$ is linearly proportional to $\rho_{xx}$, whereas, the side jump models gives $\rho_{xy}^{A}\propto\rho_{xx}^{2}$. The other is intrinsic Kaplus-Luttinger (KL) mechanism related to spin-orbit interaction of Bloch electronic bands, originally proposed by Karplus and Luttinger, which also gives $\rho_{xy}^{A}\propto\rho_{xx}^{2}$.\cite{Karplus}
Importantly, recent studies indicate that there is a intimate relation between the AHE and the Berry curvature of occupied electronic Bloch states.\cite{Jungwirth,Onoda1,Haldane}

Recently discovered topological semimetals (TSMs) are characterized by the symmetry protected bulk band crossings near the Fermi energy ($E_{F}$).\cite{Burkov2,XuG,WanX,WangZ} Classified by the degeneracy of nodes, Dirac and Weyl semimetals (WSMs) with the 4- or 2-fold degenerate Dirac or Weyl points, respectively, have been theoretically predicted and experimentally confirmed.\cite{WangZ,LiuZK1,WengH,XuSY,LvBQ}
Especially, in a magnetic WSM with broken time-reversal symmetry (TRS), Weyl node can be seen as a magnetic monopole in momentum space,\cite{Fang} around which there are significant contributions on Berry curvature.\cite{Burkov} Thus, if the position of Weyl nodes are proper, especially close to $E_{F}$, and the individual Fermi surface (FS) sheets have nonzero Chern numbers, there should be large intrinsic AHE in magnetic WSMs.\cite{Burkov}
On the other hand,
the half-metallic ferromagnets (HMFMs) can also exhibit AHE, such as ferromagnetic (FM) Heusler alloy Co$_{2}$MnGa, NiMnSb and related intermetallic compounds.\cite{Ludbrook,Otto} Different from traditional ferromagnets, the HMFMs have attracted extensive interests because of the (nearly) 100\% spin polarization of conduction electrons at $E_{F}$.\cite{de Groot} They can be perfectly applied to spintronic devices.

In this work, we study the AHE and electronic structure of HMFM Co$_{3}$Sn$_{2}$S$_{2}$ with kagome layers of Co. As previous experimental results reported on
polycrystal, the compound orders ferromagnetically at $T_{C}=$ 177 K and the spontaneous magnetic moment is about 0.3 $\mu_{B}/\rm Co$.\cite{Vaqueiro,Kubodera,Schnelle}
By employing systematic first-principles calculations, transport, and angle-resolved photoemission spectroscopy (ARPES) measurements on Co$_{3}$Sn$_{2}$S$_{2}$ single crystal, we reveal the quadratically scaling relationship between anomalous Hall resistivity $\rho_{xy}^{A}$ and $\rho_{xx}$, and the linear behavior between intrinsic anomalous Hall conductivity (AHC) and the magnetization. Furthermore, the agreement between the experimental band structures and theoretical calculations support the
existence of magnetic Weyl fermions in Co$_{3}$Sn$_{2}$S$_{2}$. We suggest the main contribution of observed large AHE originates from the intrinsic mechanism, which is intimately related to the Weyl nodes near $E_{F}$.

\section*{Results}

\textbf{Structure and Ferromagnetic state in Co$_{3}$Sn$_{2}$S$_{2}$.}
 As shown in Fig. 1(a), the crystal structure of Co$_{3}$Sn$_{2}$S$_{2}$ is composed of the slabs of CoSn$_{4}$S$_{2}$ octahedra stacking in a hexagonal packing (A-B-C fashion) along the $c$ axis. Each Co atom is surrounded by four Sn and two S atoms, forming a distorted octahedron. The CoSn$_{4}$S$_{2}$ octahedra connect each other along $ab$ plane by face-sharing and along $c$ axis by corner-sharing. On the other hand, the Co atoms form perfect kagome layer with corner-sharing triangles of Co atoms in Co-Sn layer. There are two kinds of Sn sites. A half of Sn atoms lie in the centers of the kagome hexagons (Sn2 sites) (Fig. 1(b)) and another half of Sn atoms located between the Co-Sn bilayers (Sn1 sites), connecting adjacent Co-Sn2 layers. S atoms are located below and above the Co-Sn2 layers. For the plate-like Co$_{3}$Sn$_{2}$S$_{2}$ single crystal, the $c$ axis is perpendicular to the crystal surface (Supplementary Figure S1).

The $\chi(T)$ curves with zero-field-cooling (ZFC) and field-cooling (FC) modes at $\mu_{0}H=$ 1 T for $H\Vert c$ increases rapidly when $T$ is below the Curie temperature $T_{C}$ ($\sim$ 174 K) (Fig. 1(c)),
well agree with previous results in the literatures.\cite{Vaqueiro,Kubodera,Schnelle} Moreover, the ZFC and FC $\chi(T)$ curves overlap each other very well, suggesting that magnetic domains have aligned with the direction of external field when $\mu_{0}H=$ 1 T. The field dependence of magnetization $M(\mu_{0}H)$ at $T=$ 5 K for $H\Vert c$ further confirms the ferromagnetism at low temperature (inset of Fig. 1(c)). The $M(\mu_{0}H)$ curve shows a pronounced hysteresis. The hysteresis loop is square and changes its direction at very little coercive field $\mu_{0}H_{c}$ close to 0.08 - 0.1 T. The saturated magnetization $M_{s}$ about 0.3 $\mu_{B}$/Co, consistent with previous results.\cite{Vaqueiro,Schnelle}
In contrast, when $T\gg T_{c}$, the hysteresis behavior vanishes and the $M(\mu_{0}H)$ curve only shows a paramagnetic behavior (inset of Fig. 1(c)).

As shown in Fig. 1(d), the zero-field longitudinal resistivity $\rho_{xx}(T,0)$ of Co$_{3}$Sn$_{2}$S$_{2}$ shows a metallic behavior in the whole measuring temperature range. A kink can be clearly observed in the $\rho_{xx}(T,0)$ curve around $T_{C}$. The resistivity decreases rapidly with decreasing temperature below $T_{c}$ due to the decrease of spin disorder scattering. When the external magnetic field is applied ($\mu_{0}H=$ 9 T), the metallic behavior of $\rho_{xx}(T)$ has almost no change, but the kink of $\rho_{xx}(T)$ near $T_{c}$ becomes smooth and the staring temperature where the $\rho_{xx}(T)$ curve changes the slope shifts to higher temperature.


\textbf{Theoretical electronic structure, Weyl nodes and calculated AHC.}The results of the first principle calculation confirm the half-metallic feature of Co$_{3}$Sn$_{2}$S$_{2}$ at FM state (Fig. 2(a)) and reveal that the bands dispersion near the $E_{F}$ are contributed mainly by the $3d$ orbitals of Co with the polarized magnetic momentum of 0.33 $\mu_{B}$/Co irrespective of the presence of spin-orbit coupling (SOC) or not.
Co$_{3}$Sn$_{2}$S$_{2}$ belongs to the type I$_{A}$ half metallic ferromagnets and the spins of all electrons are fully polarized along the up direction.\cite{Coey}
When the SOC is absent, the valence and conductance bands inverse near the $L$ point and form a nodal ring (signed by the dashed circles in Fig. 2(a)), which is protected by the mirror symmetry of the plane M(010) shown in Fig. 2(f). When the SOC is included, it will degenerate into a pair of Weyl points (WPs) with opposite chirality off the high-symmetry lines (Fig. 2(b)). The precise position of the WPs have been obtained by calculating the Wilson-loop evolution \cite{YuR} on a much denser $k$-grid in the BZ. There are three pairs of WPs in the BZ in total, as illustrated in Fig. 2(c) and Supplemental Table S1. From the energy dispersion along the direction connecting the WPs of W1 and W2, the splitting distance of them is about 0.3 \AA$^{-1}$ (Fig. 2(d)), which is long enough to guarantee the robustness of the WPs. We have also calculated the FS of (001) surface on the energy cut of WPs with Green function method.\cite{Sancho,WuQ} As shown in Fig. 2(c), the position of the WPs with opposite chirality are projected onto (001) surface with green and white dots respectively. Since all of the WPs are covered by bulk states, it's difficult to clarify the Fermi arc that connecting the WPs from the surface state calculation.

The existence of Weyl points implies that there would be an large intrinsic AHE in Co$_{3}$Sn$_{2}$S$_{2}$. The AHC $\sigma_{xy}$ obtained from the integral of Berry curvature along $k_{z}$ in the Brillouin zone (BZ) has a very large value of about -1310 $\Omega$ cm$^{-1}$ at 0 K (Fig. 2(e) and Supplement Fig. S3). Moreover, because the time-reversal symmetry is broken in the FM state, the AHE has a direct relation with the intrinsic magnetic momentum. 
We restricted the polarized magnetic momentum as 0.09, 0.15, 0.19 and 0.33 $\mu_{B}$/Co and as illustrated in Fig. 2(e) and Supplemental Fig. S3, the AHC decreased almost linearly with the decreasing of magnetic momentum. Meanwhile, the WPs evolve with the magnetic momentum decreases from 0.33 to 0.09 $\mu_{B}$/Co as indicated by the red arrows in Fig. 2(f). The WPs with opposite chirality will approach each other slightly when the spin polarization is reduced.

\textbf{Experimental electronic structure observed by ARPES measurements.}
In order to further uncover the electronic origin of AHE in Co$_{3}$Sn$_{2}$S$_{2}$,
we performed ARPES experiments on the (001) surface of samples. The overall band structures are summarized in Fig. 3. The experimental FSs in Fig. 3(b) are in
good agreement with the projections of calculated bulk FSs on the (001) surface in Fig. 3(c). The spindle-shaped FSs centered at $\overline{M}$ points, the
triangle-shaped FSs centered at $\overline{K}$ point, and the ring-like ones connecting the spindle-shaped FSs are all captured by ARPES. Specifically, the
agreement of the spindle-shaped FSs is significant since the projection of bands on the (001) surface that creating the WPs are located along the $\overline{\Gamma}-\overline{M}$ direction, i.e., the (001)-surface projections of WPs W1 and W2 are located along this direction as illustrated in Fig. 2(c),
although the WPs are buried into bulk continuum states at the chemical potential.

The detailed near-$E_{F}$ band structures along the $\overline{\Gamma}-\overline{K}$ and $\overline{\Gamma}-\overline{M}$ directions are presented in Figs. 3(d)-3(i).
We first focus on the results along the $\overline{K}-\overline{\Gamma}-\overline{K}$ direction [Figs. 3(d)-3(f)], the projections of calculated bulk bands along this direction in Fig. 3(f) well reproduce the experimental band features in Figs. 3(d) and 3(e), i.e. the hole bands at near half of $\overline{\Gamma K}$ corresponding
to the ring-like FSs and the electron bands [dashed curves in Figs. 3(d) and 3(e)] at $\overline{K}$ points for the triangle-like FSs, whose intensities are relatively weak due to the matrix element effect. The band structures along the $\overline{M}-\overline{\Gamma}-\overline{M}$ direction are displayed in Figs. 3(g)-3(i). The
observed band dispersions are more consistent with calculated bulk bands close to the $k_z\sim\pi$ plane, as illustrated by red curves in Fig. 3(g). Indicated by the
four blue cuts in Fig. 3(a), these bulk bands are located in the $k_x$-$k_z$ plane with $k_y$ = 0 and $k_{z}$ = 0.96, 0.92, 0.88, and 0.84$\Gamma T$ from bottom to
top, respectively, which are parallel to the $\overline{\Gamma}-\overline{M}$ direction. One can now obtain a good agreement between experiments and theories along this direction, where the electronic states at high-symmetry $k_{z}$ planes ($k_{z} = \pi$ here) have main contributions to the ARPES spectra in the frame of $k_{z}$ broadening effect.\cite{Strocov,Kumigashira,Takane}

\textbf{AHE of Co$_{3}$Sn$_{2}$S$_{2}$.}
 Next, we move to the AHE of Co$_{3}$Sn$_{2}$S$_{2}$. As shown in Fig. 4(a). the Hall resistivity $\rho_{xy}$ of Co$_{3}$Sn$_{2}$S$_{2}$ at high temperature shows the linear field dependence. When $T$ is close to $T_{c}$, the $\rho_{xy}(\mu_{0}H)$ curve starts to bend significantly at low field region. For $T\ll T_{C}$, the $\rho_{xy}(\mu_{0}H)$ curve increases steeply at low field and then becomes weakly field dependent at high field (above $\sim$ 0.1 T). The field dependence of $\rho_{xy}(\mu_{0}H)$ resembles those of $M(\mu_{0}H)$ (Fig. 4(b)), typical ones for ferromagnets. But the saturation values follow opposed temperature dependencies at $T\ll T_{C}$. When the saturation magnetization increases with decreasing temperature, the saturation value of the $\rho_{xy}(\mu_{0}H)$ decreases below $T_{C}$.
The maximum saturated value of $\rho_{xy}(\mu_{0}H)$ is $\sim$ 21 $\mu\Omega$ cm at $T=$ 140 K.

It is known that the Hall resistivity $\rho_{xy}$ in the ferromagnets arises from two parts,\cite{Chien} $\rho_{xy} = \rho_{xy}^{O} + \rho_{xy}^{A} = R_{0}B+4\pi R_{s}M$, where $\rho_{xy}^{O}$ is the normal Hall resistivity due to the Lorentz force, $\rho_{xy}^{A}$ is the anomalous Hall resistivity. $R_{0}$ is the ordinary Hall coefficient, and $R_{s}$ is the anomalous Hall coefficient. From the $\rho_{xy}(\mu_{0}H)$ and $M(\mu_{0}H)$ curves (Supplementary Figure S2), the $R_{0}$ and $R_{s}$ can be determined. The $R_{0}$ is positive at entire temperatures region (inset of Fig. 4(c)), indicating that the dominant carrier is hole-type. The apparent charge carrier density $n_{a}$ can be deduced using the relation of $n_{a}\sim -1/|e|R_{0}$ (Supplemental Fig. S4), and it reaches 2.1$\times$10$^{22}$ cm$^{-3}$ at 5 K, corresponding to about 0.7 carriers per Co$_{3}$Sn$_{2}$S$_{2}$. The $R_{s}(T)$ is also positive but the absolute values is much larger that $R_{0}$ (inset of Fig. 4(d)). It increases monotonically with increasing temperature. The obtained $R_{s}$ value is 2.7$\times$10$^{-9}$ $\Omega$ cm G$^{-1}$ at 170 K, which exhibits an enhancement of two orders of magnitude when compared with the conventional itinerant ferromagnets, such as pure Fe and Ni.\cite{Volkenshtein,Kaul}

\section*{Discussion}
The relation between log$\rho_{xy}^{A}$ and log$\rho_{xx}$ at low temperature region (Fig. 4(c)) can be fitted using the formula $\rho_{xy}^{A}\propto\rho_{xx}^{\alpha}$, which gives the scaling exponent $\alpha=$ 1.87(4). It suggests that the intrinsic KL or extrinsic side-jump mechanism are mainly responsible for the AHE in Co$_{3}$Sn$_{2}$S$_{2}$. On the other hand, the dominant mechanism of AHE can also be decided by fitting the relationship between $\rho_{xy}^{A}$ and $\rho_{xx}$ using the formula $\rho_{xy}^{A}=a(M)\rho_{xx}+b(M)\rho_{xx}^{2}$. The first term corresponds to the skew scattering contribution, while the second term represents the intrinsic or side-jump contribution.\cite{ZengC} For the skew scattering contribution $a(M)$ is usually proportional to $M$ linearly.\cite{Nozieres} And for the intrinsic contribution $b(M)=\rho_{xy}^{A}/\rho_{xx}^{2}$, which is directly related to the intrinsic AHC $\sigma_{xy,\rm{in}}^{A}=-\rho_{xy}^{A}/\rho_{xx}^{2}$.\cite{Lee} Previous study suggests that the $\sigma_{xy,\rm{in}}^{A}$ is also proportional to $M$ linearly,\cite{ZengC} thus the linear fit of $\rho_{xy}^{A}/(\rho_{xx}M)$ vs. $\rho_{xx}$ can separate the intrinsic and extrinsic contributions. The relation between $\rho_{xy}^{A}/(\rho_{xx}M)$ and $\rho_{xx}$ (Fig. 4(d)) shows the linear dependence when $\rho_{xx}\leq$ 250 $\mu\Omega$ cm.
After subtracting the skew-scattering contribution, The obtained value of $|\sigma_{xy,\rm{in}}^{A}|$ at 5 K is about 494 $\Omega^{-1}$ cm$^{-1}$, which is smaller than the theoretically predicted value ($\sim$ 1310 $\Omega^{-1}$ cm$^{-1}$) when setting the moment of Co as 0.33 $\mu_{B}$ (Fig. 2(d)). This discrepancy could be due to the uncertainty of determining sample thickness or the tilt of magnetic field away from the $c$ axis during the transport measurement.
On the other hand, the $\sigma_{xy,\rm{in}}^{A}$ exhibits the linear dependence of $M$ when $T<T_{c}$ (Fig. 4(e)), well consistent with the KL theory.\cite{Karplus}
Importantly, this result is well agreement with the theoretical predictions that the smaller moment of Co leads to the nearly linear decrease of $\sigma_{xy}^{A}$ (Fig. 2(e)).
In contrast to the $|\sigma_{xy, \rm{in}}^{A}|$, the $|\sigma_{xy, \rm{sj}}^{A}|$ for the side-jump contribution can be estimated using an expression $(e^{2}/(h a)(\varepsilon_{\rm{SO}}/E_{F})$, where $\varepsilon_{\rm{SO}}$ is the spin-orbit interaction energy.\cite{Nozieres} Using the lattice constant $a\sim V^{1/3}=$ 9.96 \AA\ and $\varepsilon_{\rm{SO}}/E_{F}\sim$ 0.01 for the metallic ferromagnets, the derived $\sigma_{xy,\rm{sj}}^{A}$ is only about 3.9 $\Omega^{-1}$ cm$^{-1}$, thus the extrinsic side-jump contribution to $\sigma_{xy}^{A}$ could be very small when compared to $|\sigma_{xy, \rm{in}}^{A}|$.
To further investigate the mechanism of AHE in Co$_{3}$Sn$_{2}$S$_{2}$, the temperature dependence of $\sigma_{xy}^{A}$ is shown in Fig. 4(f). The $|\sigma_{xy}^{A}|$ shows fairly large values at low temperature region and it is insensitive to the change of temperatures. Above 140 K, however, it decreases quickly with increasing temperature and reaches to about 150 $\Omega^{-1}$ cm$^{-1}$ at 175 K. In contrast, the scale factor $S_{H}$ ($=\mu_{0}R_{s}/\rho_{xx}^{2}=-\sigma_{xy}^{A}/M$) is almost constant when $T<T_{C}$ (inset of Fig. 4(f)). Therefore it further confirms that the temperature dependence of $\sigma_{xy}^{A}$ originates from the temperature dependence of $M$, i.e., the dramatic decrease of $|\sigma_{xy}^{A}|$ above 140 K is mainly due to the sharp decrease of $M(T)$ as the temperature approaches $T_{C}$.
Above results clearly indicate that the large intrinsic AHC is dominant in HMFM Co$_{3}$Sn$_{2}$S$_{2}$, and it is closely related to the nontrivial topology of band structure with Weyl nodes near $E_{F}$.

In summary, we have investigated the AHE and electronic structure of HMFM Co$_{3}$Sn$_{2}$S$_{2}$ single crystal. The nearly quadratic relationship between $\rho_{xy}^{A}$ and $\rho_{xx}$ indicates that the mechanism of AHE in Co$_{3}$Sn$_{2}$S$_{2}$ is dominated by the intrinsic contribution. The consistency between the experimental band structures and first-principles calculations, especially the spindle-shaped FSs along the $\overline{\Gamma}-\overline{M}$ direction, indicate that the intrinsic AHE in Co$_{3}$Sn$_{2}$S$_{2}$ originates from the existence of magnetic Weyl fermions near $E_{F}$. Moreover, the steep decrease of $\sigma_{xy}^{A}$ at high temperature can be mainly ascribed to the sharp decrease of magnetic moment, causing the change of topological properties in band structure. Current work will not only deepen our understanding on the exotic physical phenomena associated with nontrivial band topology, but also shed light on exploring novel electronic/spintronic devices based on AHE and/or half metallicity.

Note added.-After we have submitted our work, we notice that there is a preprints \cite{LiuE} reporting the study on the electronic structure and AHE of Co$_{3}$Sn$_{2}$S$_{2}$ single crystal, which shares consistent conclusions to parts of ours.

\begin{methods}

\noindent\textbf{Single crystal growth and structural characterization.} Co$_{3}$Sn$_{2}$S$_{2}$ single crystals were grown by the Sn flux.\cite{Kassem}
X-ray diffraction (XRD) patterns were measured using a Bruker D8 X-ray machine with Cu $K_{\alpha}$ radiation ($\lambda=$ 0.15418 nm) at room temperature.

\noindent\textbf{Magnetization and Transport measurements.} Magnetization and electrical transport measurements were performed in a Quantum Design MPMS3 and PPMS respectively. Both longitudinal and Hall electrical resistivity were measured using a standard four-probe method on rectangular shape single crystals with current flowing in the $ab$ plane. In order to effectively avoid the longitudinal resistivity contribution due to voltage probe misalignment, the Hall resistivity was measured by sweeping the field from -5 T to 5 T at various temperatures, and the total Hall resistivity was determined by $\rho_{xy}(\mu_{0}H)=[\rho(+\mu_{0}H)-\rho(-\mu_{0}H)]/2$, where $\rho_{xy}(\pm\mu_{0}H)$ is the transverse resistivity under a positive or negative magnetic field.

\noindent\textbf{ARPES measurements.} ARPES measurements were performed at the Dreamline beamline of the Shanghai Synchrotron Radiation Facility (SSRF) with a Scienta D80 analyzer and at the 1-cubed ARPES end-station of BESSY using a Scienta R4000 analyzer. The energy and angular resolutions were set to 15 meV and 0.2$^{\circ}$, respectively. All samples were cleaved $\emph{in situ}$ along the (001) plane and measured at $T$ = 20 K in a working vacuum better than 5$\times$10$^{-11}$ Torr.

\noindent\textbf{Theoretical Calculation.}  We have simulated the electronic structure of Co$_{3}$Sn$_{2}$S$_{2}$ with first-principle calculations using Vienna $\emph{ab initio}$ Simulation Package (VASP),\cite{Kresse} and the generalized gradient approximation (GGA) of Perdew-Burke-Ernzerhof (PBE) type exchange correlation potential \cite{Perdew} was employed. The spin-orbit coupling was taken into account in all of the calculations. The surface states and anomalous Hall conductivity were calculated based on the tight binding Hamiltonian constructed by using Wannier90 package.\cite{Mostofi} The maximally localized Wannier functions for 3$d$ orbitals on Co, $5p$ orbitals on Sn and $3p$ orbitals on S have been used as the basis of the tight binding Hamiltonian. We calculate the anomalous Hall conductivity (AHC) as the sum of Berry curvatures over all of the occupied bands,\cite{Thouless}

\begin{equation}
\sigma_{xy}=-\frac{2\pi e^{2}}{h}\int_{BZ}\frac{d^{3}\vec{k}}{(2\pi)^{3}}\sum_{n}f_{n}(\vec{k})\Omega_{n}^{z}(\vec{k})
\end{equation}

Where $f_{n}(\vec{k})$ is the Fermi-Dirac distribution function, and $n$ is the index of the occupied bands. The Berry curvature can be arisen from the Kubo-formula derivation,

\begin{equation}
\Omega_{n}^{z}(\vec{k})=-2{\rm Im}\sum_{m\neq n}\frac{\langle\Psi_{n\vec{k}}|v_{x}|\Psi_{m\vec{k}}\rangle\langle\Psi_{m\vec{k}}|v_{y}|\Psi_{n\vec{k}}\rangle}{(E_{m}(\vec{k})-E_{n}(\vec{k}))^{2}}
\end{equation}

Where $v_{x(y)}$ is the velocity operator. The intrinsic AHC is calculated with 200$\times$200$\times$200 $k$-point grid based on the tight binding Hamiltonian.

\end{methods}

\begin{addendum}

\item This work was supported by the Ministry of Science and Technology of China \\
(2016YFA0300504), the National Natural Science Foundation of China (No. 11574394, 11774423, 11774421), the Fundamental Research Funds for the Central Universities, and the Research Funds of Renmin University of China (RUC) (15XNLF06, 15XNLQ07, 17XNH055).

\item[Author contributions] H.C.L., H.M.W. and S.C.W. provided strategy and advice for the research; Q.W. performed the crystal growth, magnetization and transport measurements and fundamental data analysis; R.L. performed the ARPES measurements with the assistance of Z.H.L., M.L., Y.B.H., and D.W.S., R.L. and S.C.W. analyzed the data; Y.F.X. and H.M.W. carried out the theoretical calculation; H.C.L. wrote the manuscript based on discussion with all the authors.

\item[Supplementary Information] accompanies this paper.
\item[Competing financial interests] The authors declare that they have no competing financial interests.


\end{addendum}

\begin{figure}
  \centerline{\epsfig{figure=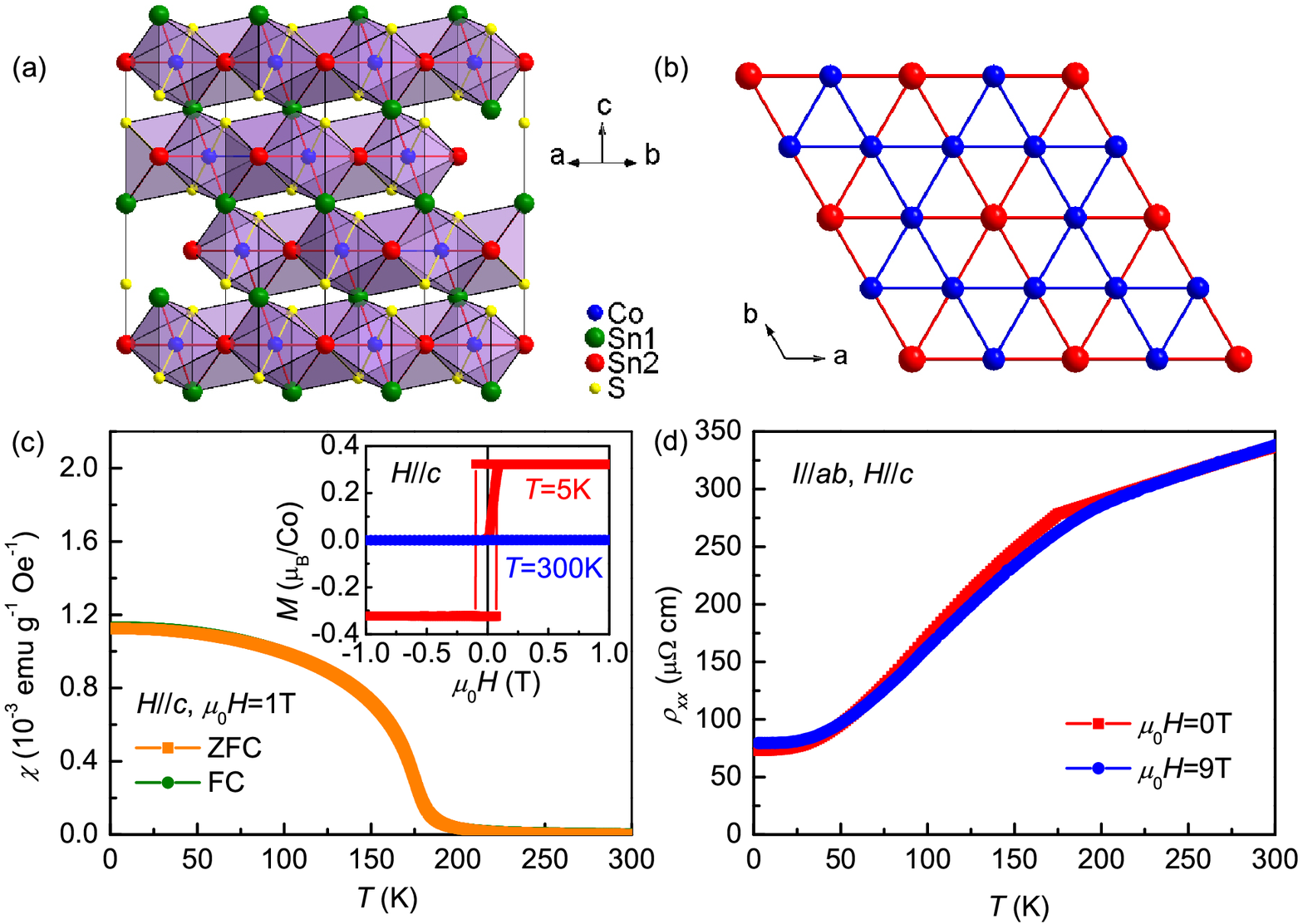,width=0.8\columnwidth}}
  \caption{\textbf{Structure, magnetization and longitudinal resistivity of Co$_{3}$Sn$_{2}$S$_{2}$.} (a) Crystal structure of Co$_{3}$Sn$_{2}$S$_{2}$ and (b) kagome layer made of Co atoms. The small blue and yellow ball represents Co and S atoms, respectively, and the big green and red ball represents Sn atoms at Sn$_{1}$ and Sn$_{2}$ sites, respectively. (c) Temperature dependence of magnetic susceptibility $\chi(T)$ with ZFC and FC modes at $\mu_{0}H=$ 1 T for $H\Vert c$. Inset: field dependence of magnetization $M(\mu_{0}H)$ at 5 K and 300 K for $H\Vert c$. (d) Longitudinal resistivity $\rho_{xx}(T,\mu_{0}H)$ as a function of temperature $T$ at $\mu_{0}H=$ 0 and 9 T along the $c$ axis.}
\end{figure}

\begin{figure}
  \centerline{\epsfig{figure=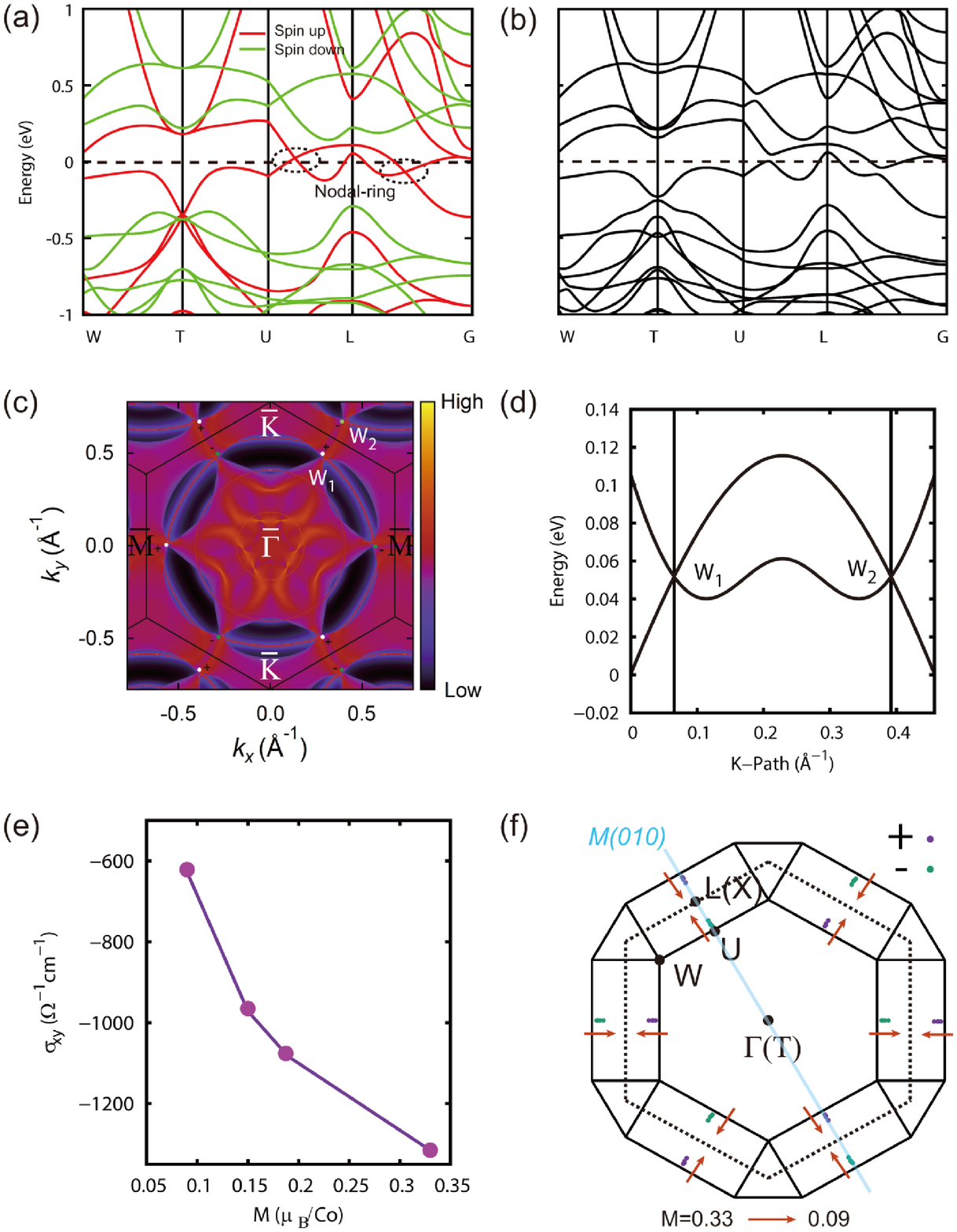,width=0.8\columnwidth}}
  \caption{\textbf{Band structure obtained from first-principle calculations.} (a) The spin-resolved band structure without SOC. (b) The band structure with SOC included. The definition of high-symmetry points are shown in Fig. 3(a). (c) Fermi surface on the (001) surface cutting on the energy of WPs. The white and green dots represent the projected WPs with positive and negative chirality respectively. (d) The energy dispersion along the direction connecting the WPs W1 and W2 in (c). (e) The intrinsic anomalous Hall conductivity with the magnetic momentum of 0.09, 0.15, 0.19 and 0.33 $\mu_{B}$/Co. (f) The evolution of WPs in the BZ with the magnetic momentum decreased from 0.33 to 0.09 $\mu_{B}$/Co. The directions of the evolution are indicated by the red arrows. The Fermi energy is set to 0.}
\end{figure}

\begin{figure}
  \centerline{\epsfig{figure=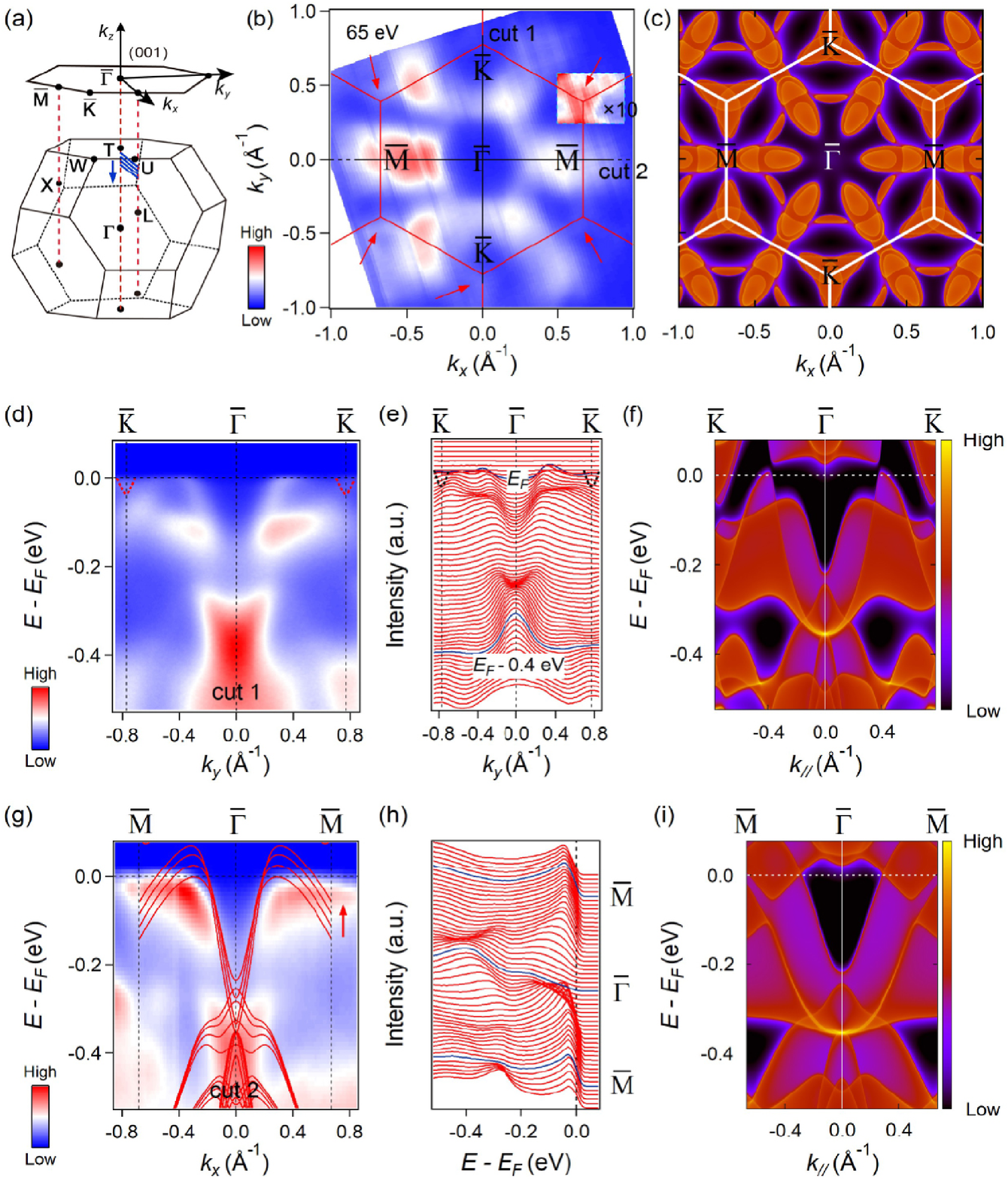,width=0.8\columnwidth}}
  \caption{\textbf{Near-$E_{F}$ ARPES spectra compared with first-principle calculations of Co$_{3}$Sn$_{2}$S$_{2}$.}
  (a) Schematic primitive BZ and 2D projected BZ of the (001) surface. Blue lines parallel to the $\overline{\Gamma}-\overline{M}$ direction
      indicate the momentum locations of calculated band structures in (g), with the $k_{z}$ momenta of 0.96, 0.92, 0.88, and 0.84$\Gamma T$
      from top to bottom, respectively.
  (b) Constant energy ARPES image obtained by integrating the spectral weight within $E_{F}$ $\pm$ 10 meV recorded with 65 eV photons. Cuts 1
      and 2 indicate the momentum locations of the experimental band structures in (d)-(i). Red lines represent the (001) surface BZs. Red arrows
      indicate the triangle-shaped FSs centered at $\overline{K}$ points in (c). The intensity within the cyan dashed rectangle has been multiplied by 10.
  (c) Projections of calculated bulk FSs on the (001) surface. White lines represent the (001) surface BZs.
  (d) and (e) ARPES intensity plot and corresponding momentum distribution curves along the $\overline{K}-\overline{\Gamma}-\overline{K}$ direction
      [cut 1 in (b)], respectively. Red and black dashed curves illuminate the electron bands forming the triangle-shaped FSs centered at $\overline{K}$
      points.
  (f) Projections of calculated bulk bands along the $\overline{K}-\overline{\Gamma}-\overline{K}$ direction.
  (g) and (h) ARPES intensity plot and corresponding energy distribution curves along the $\overline{M}-\overline{\Gamma}-\overline{M}$ direction
      [cut 2 in (b)], respectively. Red curves in (g) represent the calculated bulk bands along the four directions indicated by blue lines in (a),
      where the bands from bottom to top correspond to the cuts from top to bottom, respectively.
  (i) Projections of calculated bulk bands along the $\overline{M}-\overline{\Gamma}-\overline{M}$ direction.}
\end{figure}

\begin{figure}
  \centerline{\epsfig{figure=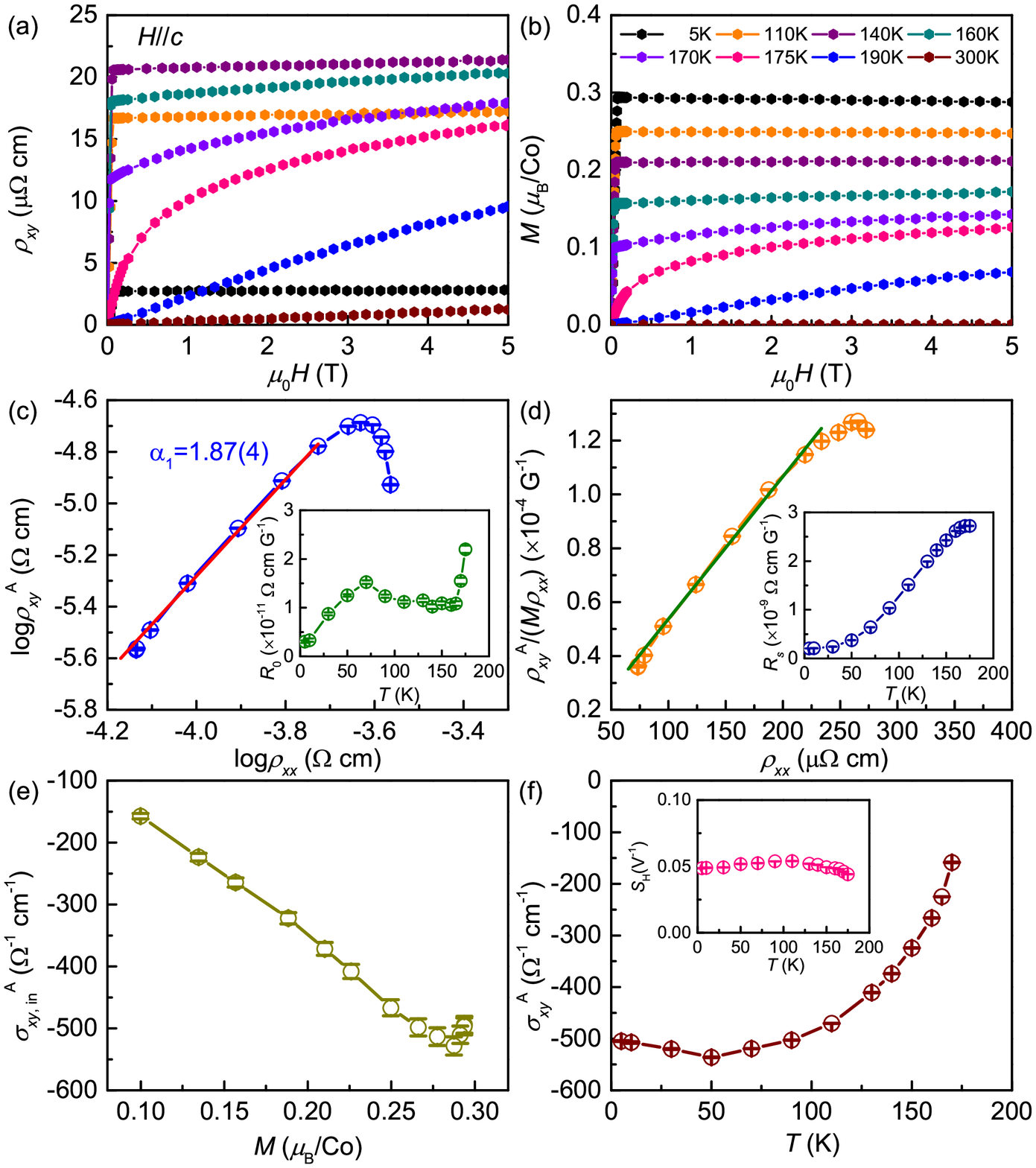,width=0.8\columnwidth}}
  \caption{\textbf{AHE of Co$_{3}$Sn$_{2}$S$_{2}$.} (a) Hall resistivity $\rho_{xy}(\mu_{0}H)$ and (b) $M(\mu_{0}H)$ as a function of magnetic field at various temperatures for $H\Vert c$. (c) Plot of log$\rho_{xy}^{A}(T)$ vs. log$\rho_{xx}(T)$. The blue solid line is the fit using the relation $\rho_{xy}^{A}\propto \rho_{xx}(T)^{\alpha}$. (d) Plot of $\rho_{xy}^{A}/(\rho_{xx}M)$ vs. $\rho_{xx}$. Insets of (c) and (d): Temperature dependence of $R_{0}(T)$ and $R_{s}(T)$ at $T\leq T_{C}$. (e) Intrinsic AHC $\sigma_{xy,\rm{in}}^{A}(T)$ as a function of $M$. (f) AHC $\sigma_{xy}^{A}(T)$ as a function of temperature. Inset: temperature dependence of scale factor $S_{H}(T)$.}
\end{figure}

\end{document}